\documentclass[twocolumn,twoside,slac_two]{revtex4}
\usepackage{graphicx}
\usepackage{fancyhdr}
\begin{document}

\title{Aspects of matter-antimatter asymmetries
in Astrophysics and relativistic
heavy ion collisions 
}

\author{F.L. Braghin}
\affiliation{ Instituto de F\'\i sica, Universidade de 
S\~ao Paulo,
 C.P. 66.318; CEP 05315-970;  S\~ao Paulo, SP, Brasil }

\begin{abstract}
Matter-anti-matter asymmetry, expected to be 
very large in the Universe, is rediscussed 
considering effects which might not have been 
considered entirely before and which can also be relevant for
(high energy densities) relativistic heavy ion collisions
Effects 
from the phase diagram of strong interactions are raised for that.
\end{abstract}

\maketitle

\thispagestyle{fancy}

\section{Introduction}

The idea of antimatter as a negative energy solution of 
relativistic free fermion equation as given by 
Dirac raises the issue of why the (observed) Universe is 
apparently constituted by matter whereas fundamental theories
of elementary particle 
indicates the same laws for (creating) matter and antimatter  
\cite{COSMO-quintess,DINEKUSENKO,QFT,COHEN-etal98,CASADEI}.
In fact, observational Astrophysics and Cosmology indicate
a large  baryon-antibaryon asymmetry in our Universe  
from cosmic background radiation (CBR) and cosmic rays observations 
\cite{DINEKUSENKO,COHEN-etal98,CASADEI}.
The most successful mechanisms which can generated this 
asymmetry are based on nonequilibrium conditions 
(together with non conservation of baryonic number, 
 CP violation) of the
early Universe, as proposed by Sakharov \cite{SAKHAROV}, 
although most of them 
do not seem to be sufficient
 \cite{DINEKUSENKO}.
The existence of 
"hidden" antimatter could  provide a reasonable solution for this 
problem 
although antimatter in cosmic rays data and other experimental data
analysed in some situations such as to yield domains ("islands")
do not provide basis to believe that galaxies/stars of antimatter
could be present neither in the closer clusters of galaxies nor 
in distances smaller than the size of the observed Universe 
\cite{COHEN-etal98}.
Photons resulting from matter-antimatter annihilation in the borders of 
the islands would be present
 in cosmic rays/CMBR and their absence can be indication of no 
 antimatter islands for regions 
smaller than nearly 20 Mpc.
Other absent observational evidences for the discredit of the existence
of antimatter islands are \cite{CASADEI}: 
the missing of (very) light (primordial) anti-nuclei
($\overline{He}$) emitted from anti-stars in cosmic rays 
and finally the non-observation of antineutrinos from antisupernovae
events - 
which would be a strong indication of the corresponding 
antimatter processes for realising energy in such
objects \cite{CASADEI}.
It is our believe however that the general idea of 
"hidden antimatter" (making the Universe less baryonic-antibaryonic asymmetric,
if not totally symmetric) 
cannot be completely discarded yet.
There are still observations to complete the present 
knowledge: searching for light antinuclei in cosmic rays,
investigating with more details the CMBR anysotropies and polarization
(Planck mission) and eventually searching antimatter 
in other forms or places 
 \cite{CASADEI}.
The breaking of CP symmetry is a relevant effect which would have allow
for the baryogenesis.
One possible effect which can be responsible for CP violation
is briefly reminded, namely, the formation of a pseudoscalar condensate,
 being amplified in a dense medium
as observed in experimental reactions. 
This mechanism of CP violation is not really completely
forbidden by fundamental theorems 
\cite{VAFA-WITTEN,IWARA,ISMD06}.
 CPT symmetry  might also have been broken and, 
in this case, astrophysical matter - antimatter asymmetry would have to be 
 reanalysed.
In this sense it is worth to remind that
CPT theorem is strictly valid for local gauge 
field theories in Minkowski spacetime, making more 
certain its behavior
in the early Universe and eventually in certain astrophysical objects.
However, in this communication, different scenarios for the 
existence of (primordial) antimatter are still considered \cite{IWARA}.

In this article the following issues are addressed 
(based in \cite{IWARA,TNewStates,SSBVAC}):
aspects of antimatter components
in Astrophysics,
eventually  considering primordial antimatter 
as hidden antimatter, which could yield
small corrections to the Hubble's law;
 associated issues of relevance for relativistic heavy ion collisions.
With experiments in (relativistic and high energy) heavy ion collisions
in BNL and CERN   the investigation of 
matter and antimatter production rates at high energy densities 
has been largely favored \cite{he-rhic}.
For these subjects, 
aspects of the phase diagram of strong interactions
with (spontaneous) symmetry breakings expected and/or envisaged 
to occur are briefly discussed.

\section{ General Aspects  }

A general field theory with fermions, vector (gauge), and spin zero fields
 ($\phi_i$, including  interacting terms $V[\phi_i]$),
 in curved space time with non minimal coupling of gauge and scalar fields
to the gravitational field can be given by \cite{BIRREL-DAVIES-FULLING}:
\begin{equation}  \label{action-gen}
\displaystyle{ 
S = \int d^4x \sqrt{-g} \left\{
\frac{i}{2} \bar{\psi} \left( \gamma_{\mu} 
{\cal D}^{\mu}-m -a_1 \Gamma_i \phi^i \right) \psi 
+ 
{\cal L}_{\phi_i,A_{\mu},R({\bf x})}
\right\} 
,
}
\end{equation}
where $\sqrt{- g}$ is the square root of the determinant of the
metric, ${\cal D}^{\mu}$ is a covariant
derivative with gauge and vector fields,
$R ({\bf x})$ is the Ricci scalar, and the various Lagrangian densities
 are denoted simply by ${\cal L}_k$.
In most part of this work it is assumed that at least one 
vector field can be treated classically -
being  eventually 
associated to a spontaneous symmetry breaking,  as a "condensed" field.
This can be considered for different phases of the early Universe. 
The non minimal coupling of vector (gauge?) fields to gravity yields
a sort of "effective mass" to them  in strong gravitational
fields which may help condensation.
In such conditions spatial anysotropies can  occur \cite{ISMD06}, 
given that the vector fields usually are associated to
gauge symmetries.
The respective inhomogeneities could manifest and have 
constraints due to the measured
CMBR and formation of large structures.

In the limit of flat space time 
the  eigenvalues of the Dirac equation for fermions and antifermions
coupled to 
a classical vector field are given by
 $E^{\pm} =  g_V V_0 \pm \sqrt{ ({\bf p} + {\bf V})^2 + (M^*)^2 },$ 
where $M^*$ takes
into account terms which modify the (anti)fermionic mass.
These solutions do not
have the symmetry of the matter-antimatter in the vacuum.
Should the vector field component $V_0$ become negative 
at zero density the
eingenvalues associated to antimatter 
are more relevant but at finite densities
things are more subtle.
This component of the classical vector field might also be due to
certain gluonic degrees of freedom in the deconfined phase of QCD
at high temperatures/energy densities \cite{ISMD06,ISMD06-others}.

Considering that due to the curved metric and/or to in medium 
effects the fermions have their wave function such that: 
  $\vec\nabla \psi \simeq (\vec{F} - i .\vec{k}) \psi$ and
$\vec{\nabla} \bar\psi \simeq (\vec{G} + i .\vec{k}) \bar\psi$ 
where $F$ and $G$ can be constants
or functions of momenta such that it is possible to define (different)
 effective  masses for fermions and antifermions in a nonhomogeneous
configuration.

However the geometry at the scale of the Universe 
is determinant in several ways.
The Dirac equation for a fermionic and an antifermionic fields 
in  a Friedmann-Robertson-Walker metric are given by:

\begin{equation} \label{equations}
\displaystyle{ 
( i \gamma_{\mu} (\nabla^{\mu} - g_v V^{\mu}) 
 + m_{\psi}  - a_1 \phi ({\bf x}) ) \psi ({\bf x}) = 0,}
\end{equation}
\begin{equation} \label{equations2}
\displaystyle{
(  i \gamma_{\mu} ( \nabla^{\mu} - g_v V^{\mu}) 
- m_{\bar{\psi}} + a_1 \phi ({\bf x}) ) \bar\psi ({\bf x}) = 0,}
\end{equation}
where both the differential operator and the Dirac matrices depend on 
the geommetry. Besides that 
the masses were consider to incorporate eventual 
effects from CPT breakdown.
The particle number in curved space time has intrinsic 
subtleties \cite{BIRREL-DAVIES-FULLING} which will not be addressed here.

The usual scenario for describing the observed Universe
with (nearly) equal quantities of matter and antimatter
( constituted by islands / domains of matter and antimatter)
considers that inflation would have kept these domains apart 
hindering mutual annihilation \cite{COHEN-etal98,CASADEI}. 
However considering that most (if not all) matter (antimatter)
from the early Universe was created just after inflation
it is reasonable to think about domains of matter and antimatter
which could have been kept apart due to particular mechanisms 
\cite{ISMD06,CASADEI}
reducing the flow of particles towards the annihilation zone \cite{COHEN-etal98}.

\subsection{ Finite densities in Minkowski space}

Asymmetries between fermions and anti-fermions at 
finite density (chemical potential) could have 
had different configurations and dynamics in the early Universe,
as well as they seem to exhibit in relativistic heavy ions collisions,
and some other examples and cases are considered elsewhere
\cite{IWARA,ISMD06,SSBVAC}.

\subsection{ Some speculative scenarios }

Different scenarios for 
the matter-antimatter (inhomogeneous) configurations can be formulated
for different matter-antimatter asymmetries.

One of the most investigated possibilities for the problem 
of antimatter in Cosmology was called
''antimatter islands'' as discussed above \cite{COHEN-etal98}. 
Different ways of explaning why the observations discussed above
do not provide  evidences for these islands is proposed in the following.

Even if the existence of stars/galaxies of antimatter would be
ruled out in the future cosmic rays 
investigations \cite{CASADEI},  other possibilities remain open.
For instance, 
these light nuclei could have been suppressed in collisions
before reaching Earth. 

However  if CPT had been 
broken in the Early Universe 
in such a way as to make antimatter domains 
to collapse faster than matter 
 "antimatter made black holes" (such  the Primordial Black Holes)
there would have been formed. 
These 
"anti-black holes" could eventually be responsible for a 
considerable amount of (hidden) antimatter.
They could be even present in very energetic places
such as the center of the Galaxy \cite{centergalaxy}.
Effects in the dynamics of black holes and anti black holes may not
 be observed however.

Finally consider the dynamics of relativistic heavy ion
collisions in particular associated to the antiparticle-particle
ratios (which increase with the increase of energy densities) 
created at finite energy density \cite{he-rhic}.
Besides the possible contribution of 
(nonhomogeneous) vector fields (classical or not)
as discussed above for the case of particle-antiparticle ratios
and configurations 
it is also argued in  \cite{SSBVAC,ISMD06} that the 
different ratios in the yields of baryons and antibaryons 
in relativistic heavy ions collisions may be a signature
of the restoration of the spontaneously breakdown of chiral symmetry.


\subsection{ Other developments}

\noindent{  Antimatter  in dense stars: di-antiquarks condensation}.
Some partial effects of classical tensor and vector fields,
eventually associated to classical gluonic configurations,
were considered to the formation of superconductive states
at very high densities in a schematic model.
These classical fields can favor the appearance of
 condensates of di-antifermions $<\overline{q} \overline{q}>$
besides the usual di-fermions (di-quarks) condensates $<q q>$
in color superconductivity 
in a way similar to that showed above for finite density fermions
\cite{FewBody2006}.
There is the
 possility of coexistence in dense stars.

\noindent{\bf Raising issues on Hubble's law}
Deviations from the Hubble's law expansion (eventually 
to be coped with ideas related to the so called "quintessence" 
\cite{COSMO-quintess})
can be suggested by:
(i) the observation of very far Supernova type I in the edges of 
the observed Universe,
(ii) the small fluctuations in different scales of distance-speed of 
recession of galaxies.
The formation of large scale structures
 from fluctuations still allows to ask  whether
Hubble's law \cite{HUBBLE}
has anisotropic corrections compatible with observations
\cite{IWARA,COSMO-quintess}.
In relativistic heavy ions collisions the flow of particles 
usually follow a quite well defined Hubble's flow.
However there still are hints of deviations \cite{ISMD06-others}.

\section{ 
 Concluding Remarks}

In this article several scenarios were discussed 
for the matter-antimatter asymmetry of the visible Universe.
Some of the aspects can be eventually useful for the 
investigation of antimatter present
in relativistic heavy ions collisions.
Some issues relevant also for the Hubble's law
(and eventually for the corresponding Hubble's flow 
in relativistic heavy ions collisions)  were also briefly
discussed.
It was pointed out that CPT invariance might have been 
broken (more probably in the past), besides the usual breakdown of 
T reversal (and the corresponding CP invariance) which 
is observed in strong and weak processes, contributing either
to the stronger baryogenesis or the different scenarios in which 
the baryon-antibaryon asymmetry is smaller than 
observed nowadays in  Astrophysics. 
This issue can have also relevance for the understanding of other deep
questions such as: where, when, how and at which level does time arrow 
appear
 such that (our) "thermodynamical" Universe emerges? 
would an 
"antimatter thermodynamical domain of Universe" have the same behavior and laws
of our matter dominated domain (if it is really a domain)
In this sense it seems to be fair to ask whether
light antinuclei ($He$) 
from anti-stars could be expected to be as abundant as the light nuclei
from stars ?

\bigskip 
\begin{acknowledgments}
This work was supported in part by FAPESP. 
The author thanks 
the organizing committee for the
nice workshop.
\end{acknowledgments}

\end{document}